\documentclass[twocolumn,showpacs,preprintnumbers,amsmath,amssymb,pra]{revtex4-1}

\usepackage{graphicx}
\usepackage{dcolumn}
\usepackage{bm}

\usepackage[T1]{fontenc} 

\newcommand{\nonum}{\item[]}

\newcommand{\etal}{\textit{et al$\ $}}

\newcommand{\Name}[1]{#1}
\newcommand{\REVIEW}[4]{{#3}\ {#1}\ \textbf{#2}\ {#4}}

\usepackage{color}
\newcommand{\wyroz}[1]{{#1}}%
\newcommand{\wyr}[1]{{#1}}%

\begin{document}

\preprint{Submitted to: JOURNAL OF PHYSICS: CONDENSED MATTER}

\title{The magnetic field induced phase separation  \\ in a~model of a superconductor with local electron pairing }

\author{Konrad Kapcia}%
    \email{corresponding author; e-mail: kakonrad@amu.edu.pl}
\author{Stanis\l{}aw Robaszkiewicz}%
\affiliation{Electron States of Solids Division, Faculty of Physics, Adam Mickiewicz University in Pozna\'n, Umultowska 85, PL-61-614 Pozna\'n, Poland, EU
}%

\date{December 23, 2012}

\begin{abstract}
We have studied the extended Hubbard model with pair hopping in the atomic limit for arbitrary electron density and chemical potential and focus on paramagnetic effects of the external magnetic field.
The Hamiltonian considered consists of (i)~the effective on-site interaction $U$ and (ii)~the intersite charge exchange interactions $I$, determining the hopping of electron pairs between nearest-neighbour sites.
The phase diagrams and thermodynamic properties of this model have been determined within the variational approach (VA), which treats the on-site interaction term exactly and the intersite interactions within the mean-field approximation.
Our investigation of the general case shows that the system can exhibit not only the homogeneous phases: superconducting (SS) and nonordered (NO), but also the phase separated states (PS: \mbox{SS--NO}).
Depending on the values of interaction parameters, the PS state can occur in higher fields than the SS phase (field-induced PS). Some ground state results beyond the VA are also presented.
\end{abstract}

\maketitle


\section{Introduction}\label{sec:intro}

Recently, there has been much interest in superconductivity with very short coherence length. This interest is due to its possible relevance to high temperature superconductors (the cuprates, doped bismuthates, iron-based system, fullerenes, etc.;
for a review, see \cite{MRR1990,AAS2010,RP1993,KRM2012,DHM2001} and references therein).
Also the phase separation (PS) phenomenon involving superconducting (superfluid) states is very current topic, because it can play crucial role determining behaviour in many real compounds~\cite{KPK2004,CSP2008,TCS2008,UAC2009,SFG2002,MWB2006,PIN2009,RPC2011,V1988,TKP1995} and fermions on optical lattices \cite{LH2012,CX2010,ZAS2006,PLK2006,SSS2008}.

In present work we will study paramagnetic effects of magnetic field (Zeeman term) in a~model which is a~generalization of the standard model of a local pair superconductor with on-site pairing (i.~e. the model of hard core bosons on a~lattice \cite{MRR1990,BBM2002,MRK1995}) to the case of finite pair binding energy.
Such analysis of paramagnetic effects is important, in particular, for unconventional superconductors, for which the temperature dependence of the upper critical field has positive curvature \cite{OSW1993,MJL1993} and does not saturate even at genuinely low temperature \cite{MJL1993}. Also the pseudogap is  destroyed by sufficiently high field \cite{SKM2001}.
Moreover, recently the possibility of the magnetic field induced phase separation (SS/NO) has been found for the $d=2$ and $d=3$ dimensional spin-polarized attractive Hubbard model \cite{KM2011} as well as for the continuum fermion model in $d=2$ \cite{LZ2008}.

The Hamiltonian considered has the following form:
\begin{equation}\label{row:ham1}
\hat{H}=U\sum_{i}{\hat{n}_{i\uparrow}\hat{n}_{i\downarrow}}- 2I\sum_{\langle i,j\rangle}{\hat{\rho}_i^+\hat{\rho}_j^-} - \mu\sum_i\hat{n}_i - B\sum_i{\hat{s}^z_i},
\end{equation}
where  \mbox{$\hat{n}_{i}=\sum_{\sigma}{\hat{n}_{i\sigma}}$}, \mbox{$\hat{n}_{i\sigma}=\hat{c}^{+}_{i\sigma}\hat{c}_{i\sigma}$}, \mbox{$\hat{\rho}^+_i=(\hat{\rho}^-_i)^\dag=\hat{c}^+_{i\uparrow}\hat{c}^+_{i\downarrow}$}. $B=g\mu_BH_z$ is external magnetic field and \mbox{$\hat{s}^z_i=(1/2)(\hat{n}_{i\uparrow}-\hat{n}_{i\downarrow})$} is $z$-component of the total spin at $i$ site.
$\hat{c}^{+}_{i\sigma}$ denotes the creation operator of an electron with spin \mbox{$\sigma=\uparrow,\downarrow$} at the site $i$,
which satisfy canonical anticommutation relations:
\mbox{$\{ \hat{c}_{i\sigma}, \hat{c}^+_{j\sigma'}\} = \delta_{ij}\delta_{\sigma\sigma'}$},
\mbox{$\{ \hat{c}_{i\sigma}, \hat{c}_{j\sigma'}\} = \{ \hat{c}^+_{i\sigma}, \hat{c}^+_{j\sigma'}\} = 0$},
where $\delta_{ij}$ is the Kronecker delta.
\mbox{$\sum_{\langle i,j\rangle}$} indicates the sum over nearest-neighbour sites $i$ and $j$ independently.
$U$ is the on-site density interaction,
$I$ is the intersite  charge  exchange interaction between nearest neighbours.
$\mu$ is the chemical potential, connected with the concentration
of electrons by the formula:
\mbox{$n = (1/N)\sum_{i}{\left\langle \hat{n}_{i} \right\rangle}$},
with \mbox{$0\leq n \leq 2$} and $N$ is the total number of lattice sites.

In this paper, we treat the parameters $U$ and $I$ as the effective ones, assuming that they include all the possible contributions and renormalizations like those coming from the strong electron-phonon coupling or from the coupling between electrons and other electronic subsystems in solid or chemical complexes \cite{MRR1990}. In such a general case arbitrary values and signs of $U$ and $I$ are important to consider.
\wyroz{
Formally, $I$ is one of the off-diagonal terms of the Coulomb interaction \mbox{$I_{ij}=-(1/2)(ii|e^2/r|jj)$}~\cite{H1963,H1991}, describing a part of the so-called bond-charge interaction, and the sign of the \textit{Coulomb-driven} charge exchange is typically negative (repulsive, \mbox{$I<0$}). However, the effective attractive interaction of this form (\mbox{$I>0$}) is also possible~\cite{FH1983,RMR1987,BL1988}. In \wyr{particular, it} can originate from the coupling of electrons with intersite (intermolecular) vibrations via modulation of the hopping integral~\cite{FH1983}, or from the on-site hybridization term in generalized periodic Anderson model~\cite{RMR1987,BL1988}.
}

In the analysis of model (\ref{row:ham1}) at \mbox{$T\geq 0$} we have adopted a variational approach (VA) which treats the on-site interaction $U$ exactly and the intersite interactions $I$ within the mean-field approximation \cite{RP1993,KRM2012,R1994}. Moreover, at \mbox{$T=0$} exact results for \mbox{$d=1$} and results obtained within random phase approximation (RPA) (for \mbox{$d=2$} and \mbox{$d=3$} lattices) are presented.

Model~(\ref{row:ham1}) has been investigated intensively in the absence of the external magnetic field only.
The first analysis  of the phase diagram of  model (\ref{row:ham1}) have been performed by Bari~\cite{B1973} and Ho and Barry~\cite{HB1977} using the variational method in order to examine the instability of the Mott insulator to superconductivity mostly for the special case of the half-filled band (\mbox{$n=1$}). The effects of diagonal disorder  on the critical temperature for \mbox{$U=0$} and \mbox{$n=1$} have been also determined~\cite{WA1987}, arriving at a~satisfactory qualitative interpretation of quite a number of different experiments in amorphous superconductors.
Within the VA the phase diagrams of model (\ref{row:ham1}) as a function of the electron concentration $n$ for \mbox{$B=0$} have been investigated in \cite{RP1993,KRM2012,R1994}. The stability conditions of states with phase separation for \mbox{$B=0$} have been discussed in \cite{KRM2012} only.

In this paper, we investigate model Hamiltonian (\ref{row:ham1}) for  arbitrary $\mu$ and arbitrary $n$ at \mbox{$T=0$} and finite temperatures. We focus on the effects of external magnetic field in the system.
Our investigation of the general case finds that, depending on the values of the interaction parameters and the electron concentration, the system can exhibit homogeneous SS and NO  phases as well as the PS between them. Transitions between various states and phases can be continuous and discontinuous, what implies existence of tricritical points on the phase diagrams. We present detailed results concerning the evolution of the diagrams  as a function of external field $B$, interaction parameters, $\mu$ and $n$ \wyroz{and discuss representative thermodynamic properties of the system}.
The results obtained in this work can be useful for the description of systems with local pairing in the magnetic field. They are also important as a~test and a~starting point for a perturbation expansion in powers of the hopping $t_{ij}$ and as a benchmark for various approximate approaches analyzing the corresponding finite bandwidth models.

The paper is organized as follows. In section~\ref{sec:method} we describe the VA method. Section~\ref{sec:diagrams} is devoted to the study of the phase diagrams: section.~\ref{sec:gs} includes results at \mbox{$T=0$} (VA, exact, RPA) whereas in section~\ref{sec:finitetemp} the VA results at \mbox{$T>0$} are presented.
\wyroz{In section~\ref{sec:termo} representative thermodynamic characteristics are evaluated and discussed.}
Finally, section~\ref{sec:conclusions} contains conclusions and supplementary discussion.

\section{The variational method}\label{sec:method}

Within the VA  the on-site interaction term is treated exactly and the intersite interactions are decoupled within the mean-field approximation (site-dependent):
\begin{equation}
\hat{\rho}^+_{i}\hat{\rho}^-_{j} \rightarrow \left\langle \hat{\rho}^+_{i} \right\rangle \hat{\rho}^-_{j} + \left\langle \hat{\rho}^-_{j}\right\rangle \hat{\rho}^+_{i} - \left\langle \hat{\rho}^+_{i}\right\rangle \left\langle \hat{\rho}^-_{j}\right\rangle.
\end{equation}
A~variational Hamiltonian has the following form:
\begin{eqnarray}
\hat{H}_{0}& = &\sum_i{\left[U\hat{n}_{i\uparrow}\hat{n}_{i\downarrow}-\mu \hat{n}_i -2\chi^*_i\hat{\rho}^-_i -2\chi_i\hat{\rho}^+_i \right. }+ \nonumber \\
& + & \left. \chi^*_i\Delta_i + \chi_i\Delta_i^* - B\hat{s}^z_i\right],
\end{eqnarray}
where \mbox{$\chi_i=\sum_{j\neq i}I_{ij}\Delta_j$}, \mbox{$\Delta_i^*=\langle\hat{\rho}^+_i\rangle$} and \mbox{$n_i=\langle\hat{n}_i\rangle$}.
$\hat{H}_{0}$ can be diagonalized easily
and a~general expression for the grand potential $\Omega$ in the grand canonical ensemble in the VA is
\begin{equation*}
\Omega= -\frac{1}{\beta}\ln\left\{\textrm{Tr}\left[\exp(-\beta\hat{H_{0}})\right]\right\},
\end{equation*}
where \mbox{$\beta=1/(k_{B}T)$}. The average value of operator $\hat{A}$ is defined as
\begin{equation*}
\langle\hat{A}\rangle = \frac{\textrm{Tr}\left[\exp(-\beta\hat{H}_{0})\hat{A}\right]}{\textrm{Tr}\left[\exp(-\beta\hat{H}_{0})\right]}.
\end{equation*}
$\textrm{Tr} \hat{B}$ means a~trace of operator $\hat{B}$ calculated in the Fock space.

Assuming no spatial variations of the order parameter
the grand potential per site obtained in the VA for model (\ref{row:ham1})
is given by:
\begin{equation}\label{row:grandpotential}
\omega(\bar{\mu}) = \Omega/N =  -\bar{\mu} + 2I_0|\Delta|^2 - \beta^{-1}\ln( 2 Z),
\end{equation}
where
\begin{equation*}
Z = \cosh\left(\beta \sqrt{\bar{\mu}^2+ 4|I_0\Delta|^2} \right) + \exp\left( \beta U /2 \right)\cosh{\left(\beta B/2 \right)},
\end{equation*}
\mbox{$\bar{\mu}  =  \mu - U/2$}, \mbox{$I_0 = zI$}, and \mbox{$\Delta^*  = (1/N)\sum_i{\langle \hat{\rho}^+_i\rangle}$}. $z$ denotes the number of nearest neighbours.
The free energy per site \mbox{$f=\omega+\mu n$} is derived as
\begin{equation}\label{row:freeenergy}
f(n) =\bar{\mu}(n-1) + (U/2)n + 2I_0|\Delta|^2 - \beta^{-1}\ln (2 Z).
\end{equation}

In the absence of the field conjugated with the SS order parameter ($\Delta$)  there is a symmetry between \mbox{$I>0$} ($s$-pairing) and \mbox{$I<0$} ($\eta$-pairing, $\eta$S, \mbox{$\Delta_{\eta S} = \frac{1}{N}\sum_i{\exp{(i\vec{Q}\cdot\vec{R}_i)}\langle \hat{\rho}^-_i\rangle} $}, $\vec{Q}$ is half of the smallest reciprocal lattice vector) cases for model (\ref{row:ham1}), \wyroz{which neglects single particle hopping (\mbox{$t_{ij}=0$}),} thus we restrict ourselves  to the \mbox{$I>0$} case only.

\begin{figure*}
        \centering
        \includegraphics[width=0.4\textwidth]{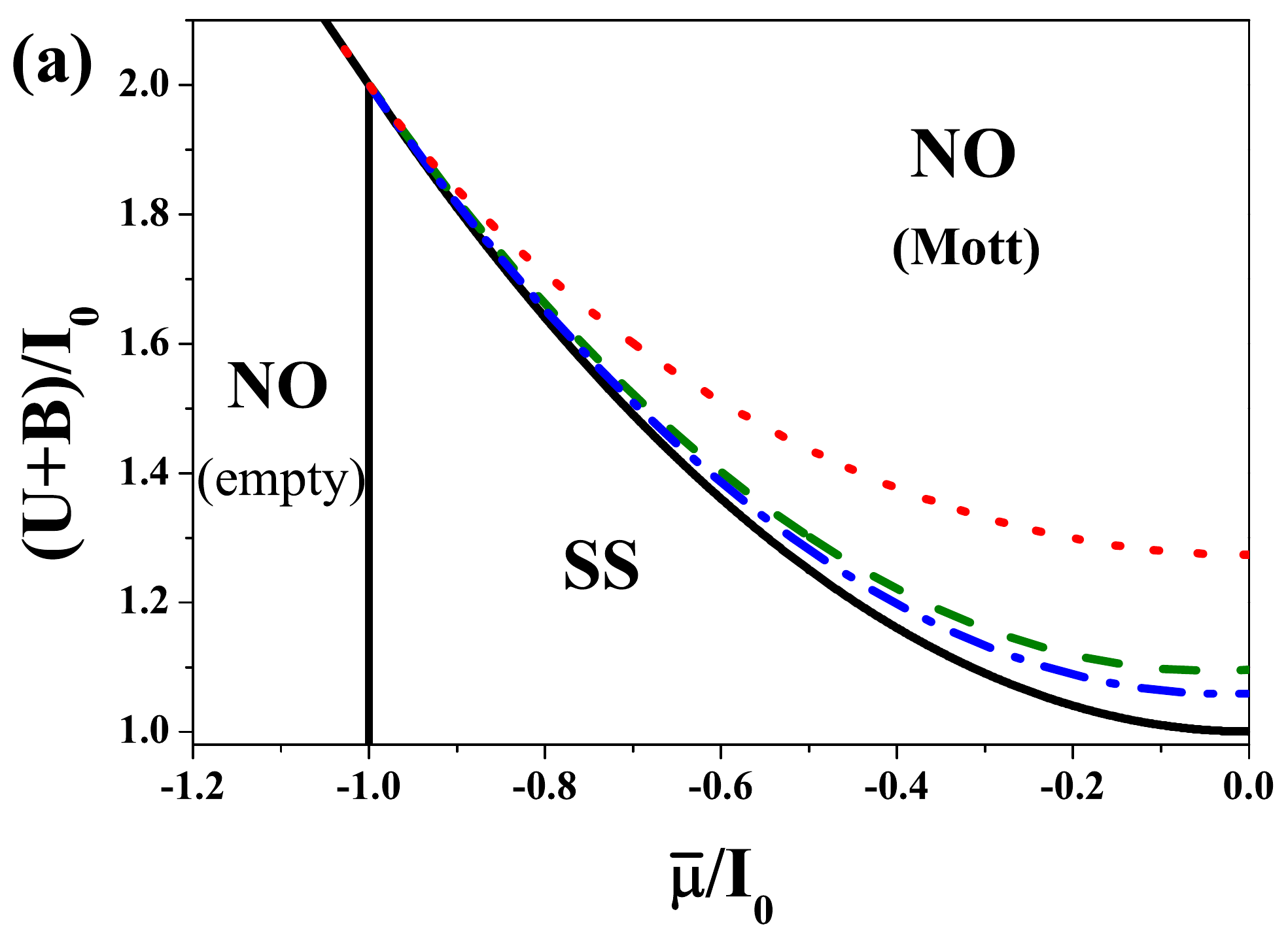}
        \includegraphics[width=0.4\textwidth]{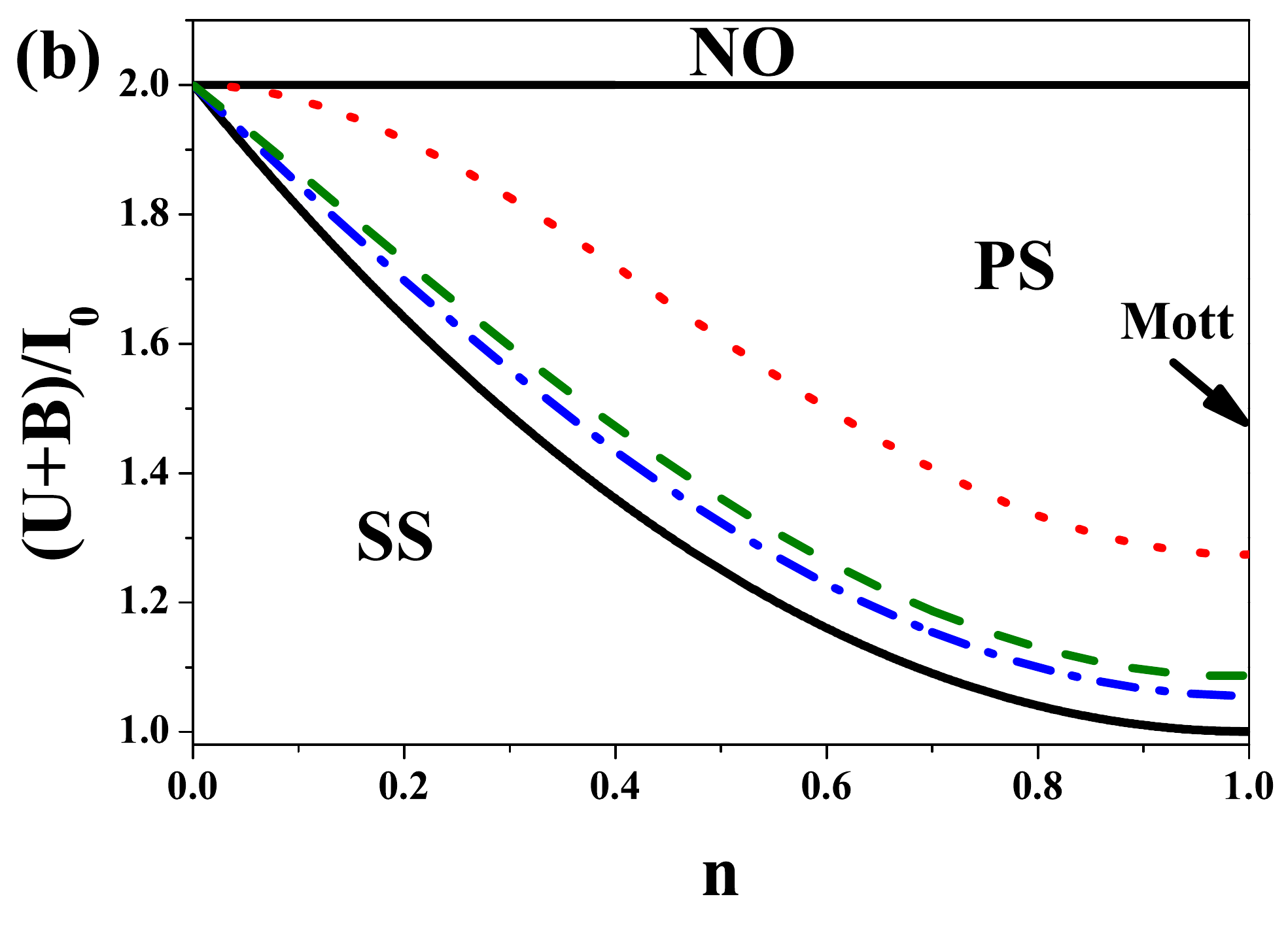}
        \caption{Phase diagrams at \mbox{$T=0$}: (a)~as a~function of $\bar{\mu}/I_0$, and (b)~as a~function of $n$. Solid, dotted, dashed, and dashed-dotted lines denote the boundaries derived within VA (exact for \mbox{$d\rightarrow+\infty$}), \mbox{$d=1$} (rigorous results), \mbox{$d=2$} (RPA, SQ lattice), and \mbox{$d=3$} (RPA, SC lattice), respectively. At half-filling (\mbox{$n=1$}) the NO (Mott state) is stable above the end of \mbox{PS--SS} lines. The \mbox{NO--SS} transition for \mbox{$\bar{\mu}/I_0=-1$} \wyr{(panel (a))} is second order, all other transitions between homogeneous phases are first order.
        The transitions to the PS state are ``third order''.%
        }
        \label{rys:GSdiagrams}
\end{figure*}

The condition for electron concentration and a~minimization of $\omega$ (or $f$) with respect to the superconducting order parameter $|\Delta|$ lead to the following self-consistent equations (for homogeneous phases):
\begin{eqnarray}
\label{row:MFA1}
\frac{\bar{\mu}\sinh\left( \beta \sqrt{\bar{\mu}^2 + 4 |I_0\Delta|^2} \right)}{Z\sqrt{\bar{\mu}^2 + 4 |I_0\Delta|^2}} = n-1, \\
\label{row:MFA2}
|\Delta|\left[ \frac{1}{I_0} - \frac{\sinh\left( \beta \sqrt{\bar{\mu}^2 + 4 |I_0\Delta|^2} \right)}{Z\sqrt{\bar{\mu}^2 + 4 |I_0 \Delta|^2}} \right]  =  0.
\end{eqnarray}
Equations (\ref{row:MFA1})--(\ref{row:MFA2})   are solved numerically for \mbox{$T\geq0$} and we obtain $|\Delta|$
and $n$ when $\mu$ is fixed or $|\Delta|$ and $\mu$ when $n$ is fixed.
The superconducting phase (SS) is characterized by non-zero value of $|\Delta|$, whereas \mbox{$|\Delta| =0$} in the non-ordered (normal) phase (NO).

The magnetization of homogeneous phases can be simply derived as
\begin{equation}
m = \langle \hat{s}^z_i \rangle = - \frac{\partial \Omega}{\partial B} = \frac{1}{2Z}\exp(\beta U/2) \sinh(\beta B/2 ).
\end{equation}
It implies that the magnetization in both SS and NO phases is nonzero  for any \mbox{$B\neq0$} and \mbox{$T>0$}.
\wyroz{In the homogeneous phases the double occupancy per site, defined as
\mbox{$D = \frac{1}{N}\sum_{i}\left\langle \hat{n}_{i\uparrow}\hat{n}_{i\downarrow}\right\rangle$},  has within VA the following form:
\begin{equation}
D = \frac{n}{2}\left[ 1 - \frac{1}{nZ}\exp{(\beta U/2)}\right].
\end{equation}
We also introduce the concentration of locally paired electrons \mbox{$n_p=2D$} and the ratio \mbox{$n_p/n=2D/n$}. Notice that $D$ is different from the condensate density (a~fraction of pairs in the condensate) \mbox{$n_0=|\langle \hat{\rho}^+ \rangle |^2$}.}

Phase separation  is a state in which two domains with different electron concentration: $n_+$ and $n_-$ exist in the system
(coexistence of two homogeneous phases). The free energies of the PS states are calculated in a~standard way, using Maxwell's construction (e.~g. \cite{KRM2012,KR2011}). In the model considered only one type of PS states can occur, which is a coexistence of SS and NO phases.

In the paper we have used the following convention. A~second (first) order transition is a~transition between homogeneous phases  with a~(dis-)continuous change of the order parameter at the transition temperature. A~transition between homogeneous phase and the PS state as a function of $n$ is symbolically named as a~``third order'' transition~\wyr{\cite{KRM2012,KR2011,KKR2010}}. \wyroz{This denotation should not be misled with the Ehrenfest's notation of order of transitions between homogeneous phases. At this transition a~size of one domain in the PS state decreases continuously to zero at the~transition temperature. One should notice that the order parameter for ``third order'' transitions is the concentration difference \mbox{$n_+-n_-$} (not $\Delta$, which is the order parameter in one domain) and its change is discontinuous at transition temperature. Such transitions are present if the system is considered for fixed $n$ and they are associated with first order transitions at fixed $\bar{\mu}$.}

All phase transition boundaries, necessary to construct the complete phase diagram within VA, have been obtained numerically
by self-consistent solving of  \mbox{(\ref{row:MFA1})--(\ref{row:MFA2})} and comparing grand potentials $\omega$ of homogeneous phases (if $\bar{\mu}$ is fixed), or free energies $f$ -- including energies of PS states -- if $n$ is fixed.

\section{Phase diagrams}\label{sec:diagrams}

The diagrams obtained are symmetric with respect to half-filling (\mbox{$n=1$}) because of the particle-hole symmetry of Hamiltonian (\ref{row:ham1}), so they will be presented only in the range \mbox{$\bar{\mu} \leq 0$} and \mbox{$0\leq n\leq 1$}.

\subsection{The ground state}\label{sec:gs}

\begin{figure*}
    \centering
    \includegraphics[width=0.325\textwidth]{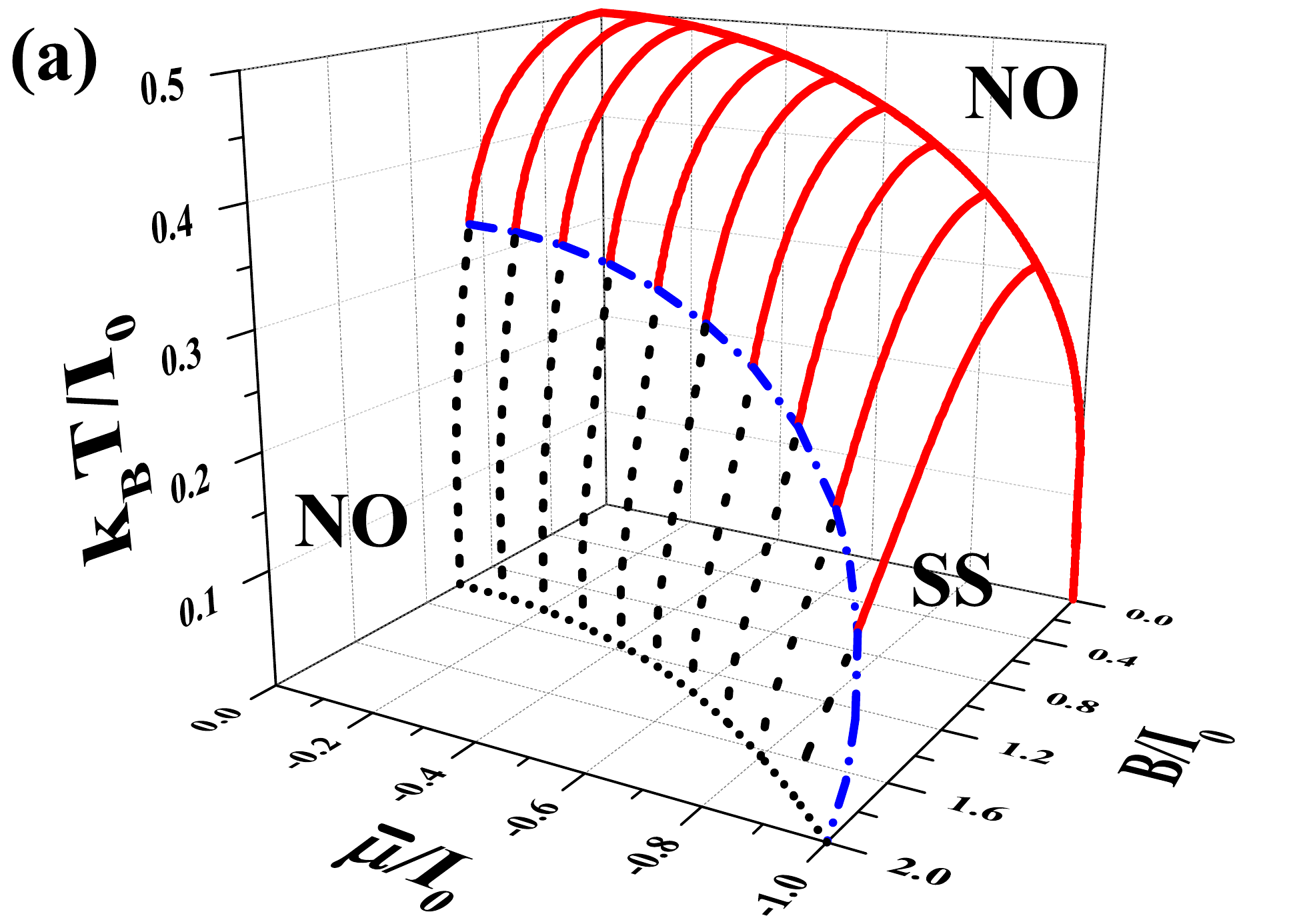}
    \includegraphics[width=0.325\textwidth]{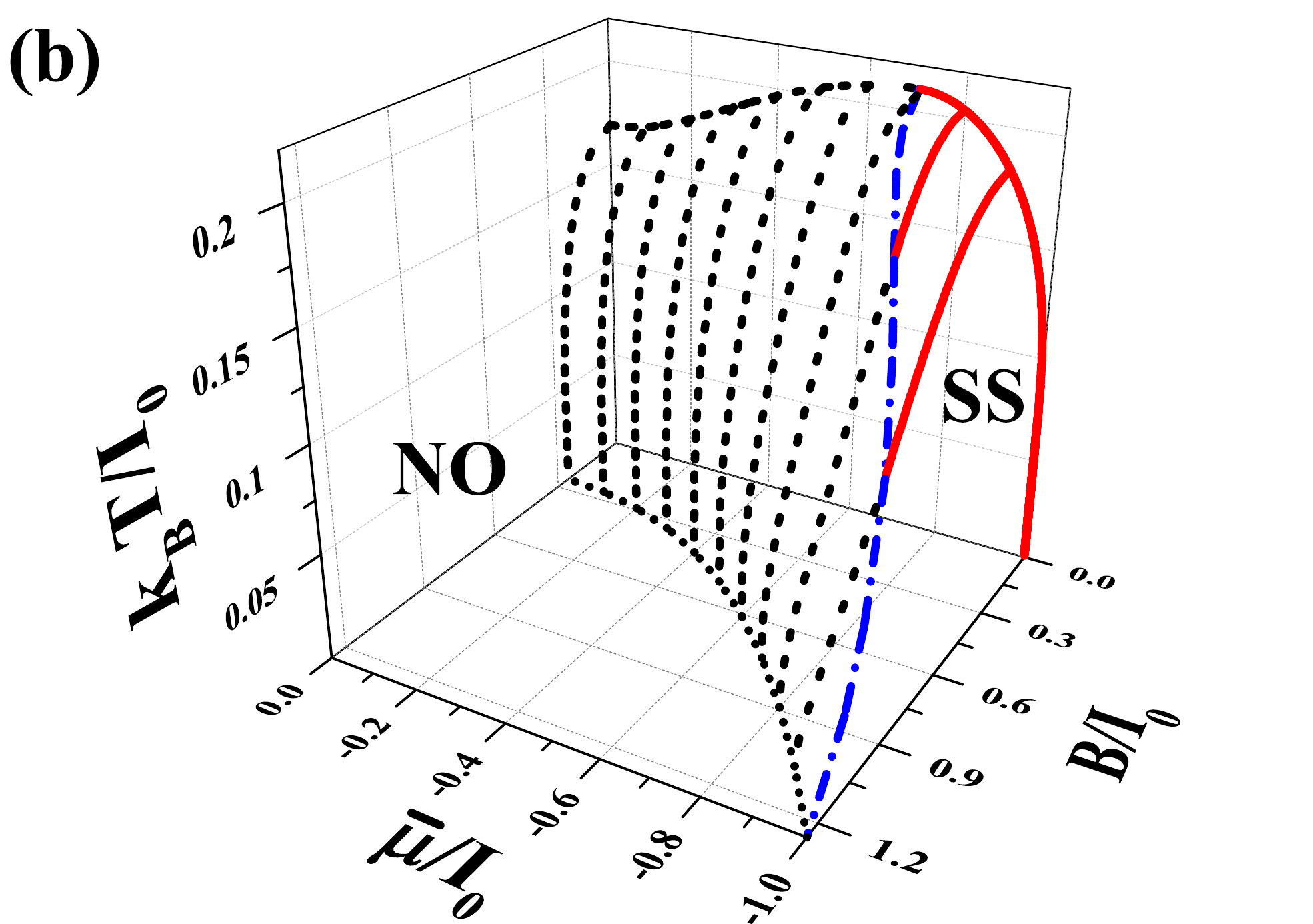}
    \includegraphics[width=0.325\textwidth]{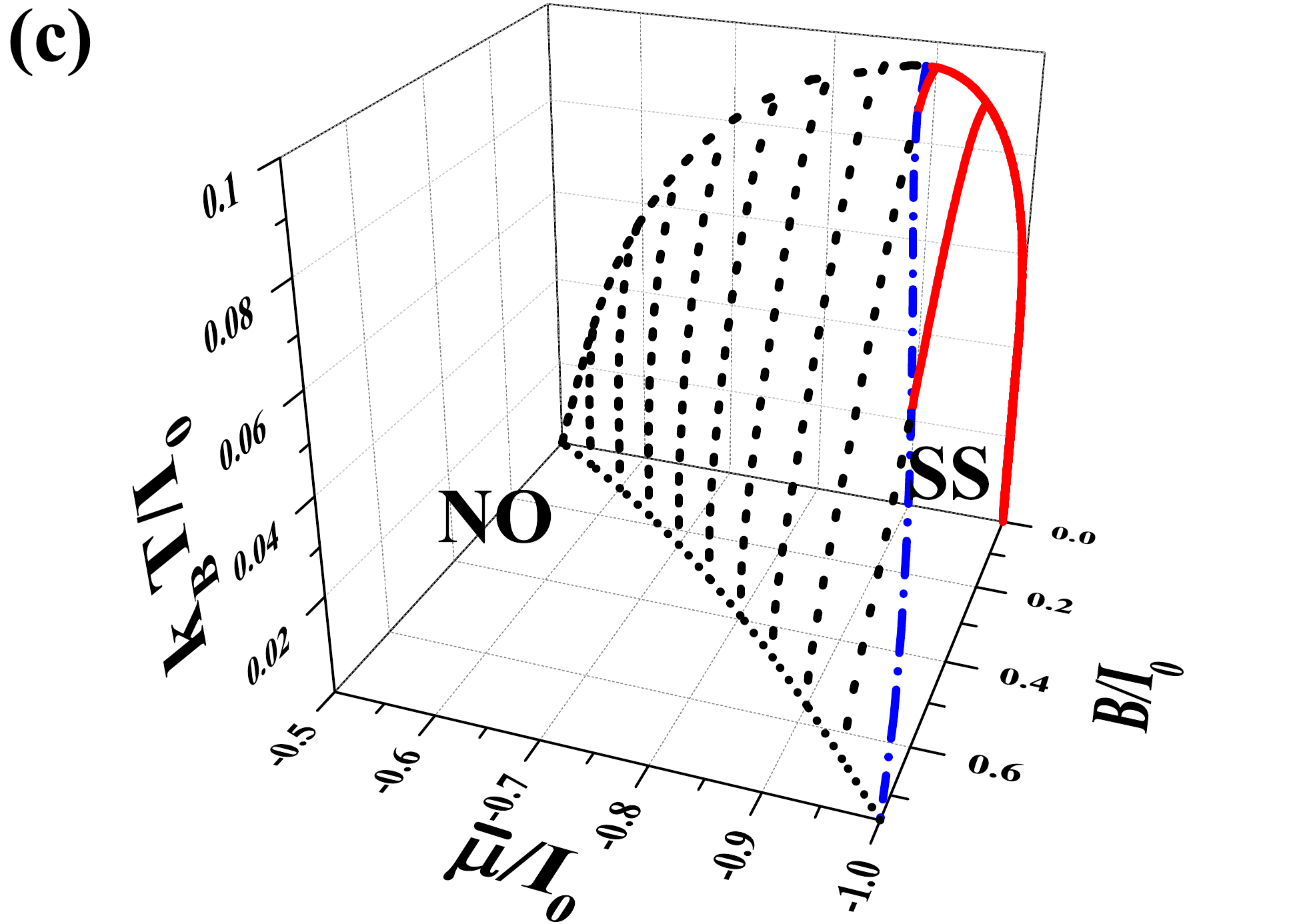}
    \caption{Finite temperature  phase diagrams for \mbox{$U/I_0=0$} (a), \mbox{$U/I_0=0.75$} (b), and \mbox{$U/I_0=1.25$} (c) plotted as a function of $\bar{\mu}/I_0$  and $B/I_0$. Solid and dotted lines indicate second order and first order transitions, respectively. The tricritical point line is denoted by dashed-dotted line.}
    \label{rys:kTvsmi3D}
\end{figure*}

\begin{figure*}
    \centering
    \includegraphics[width=0.325\textwidth]{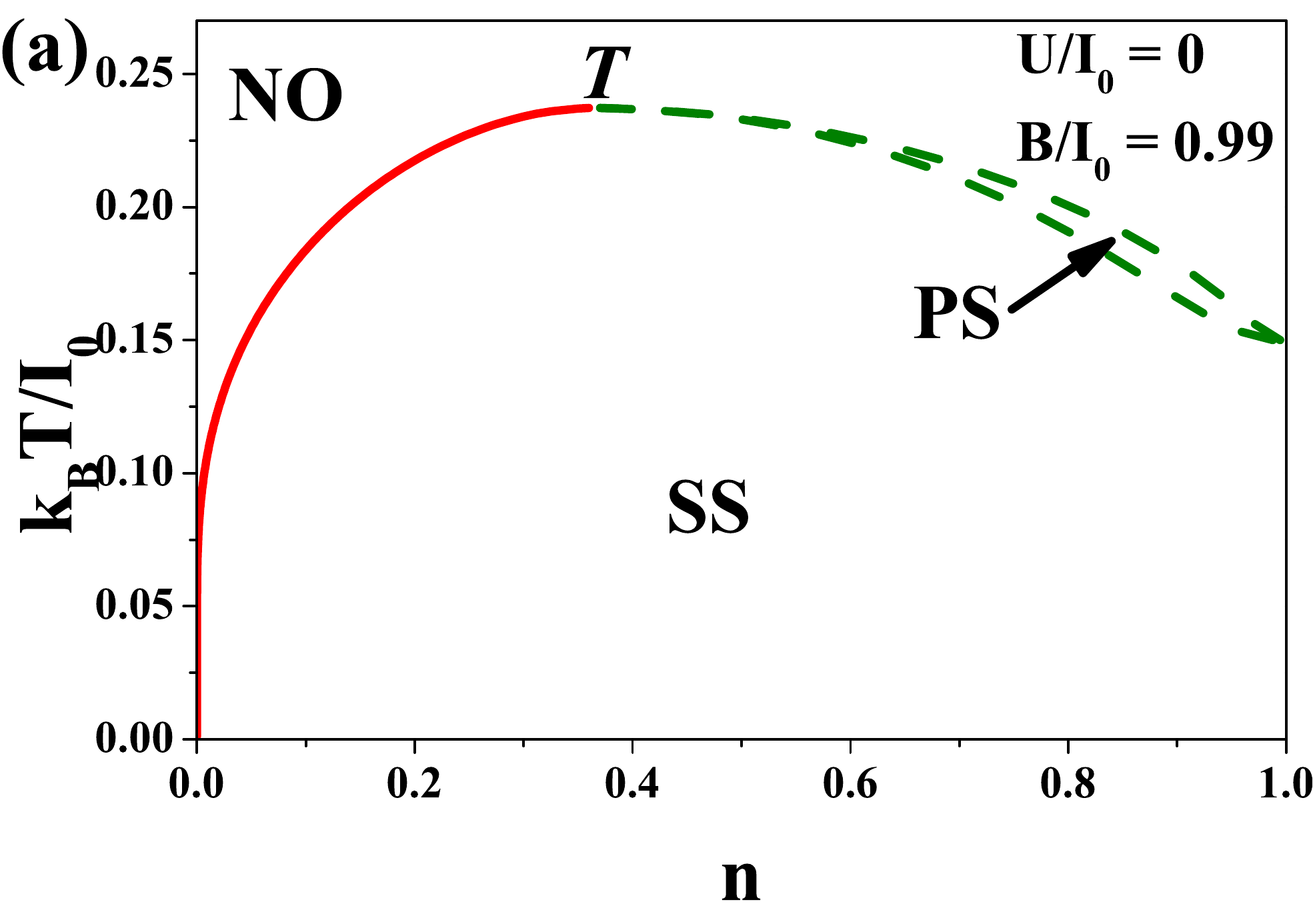}
        \includegraphics[width=0.325\textwidth]{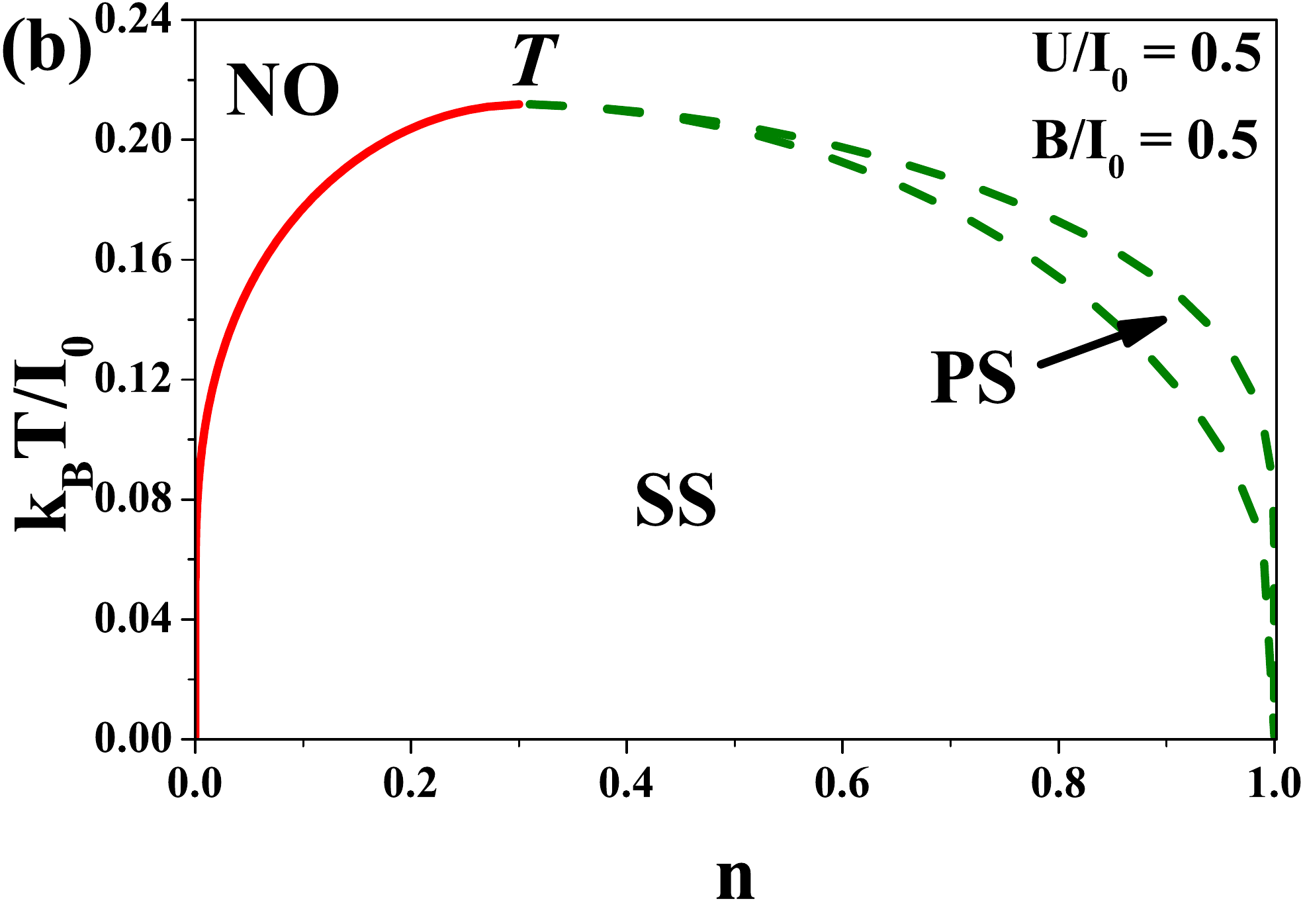}
            \includegraphics[width=0.325\textwidth]{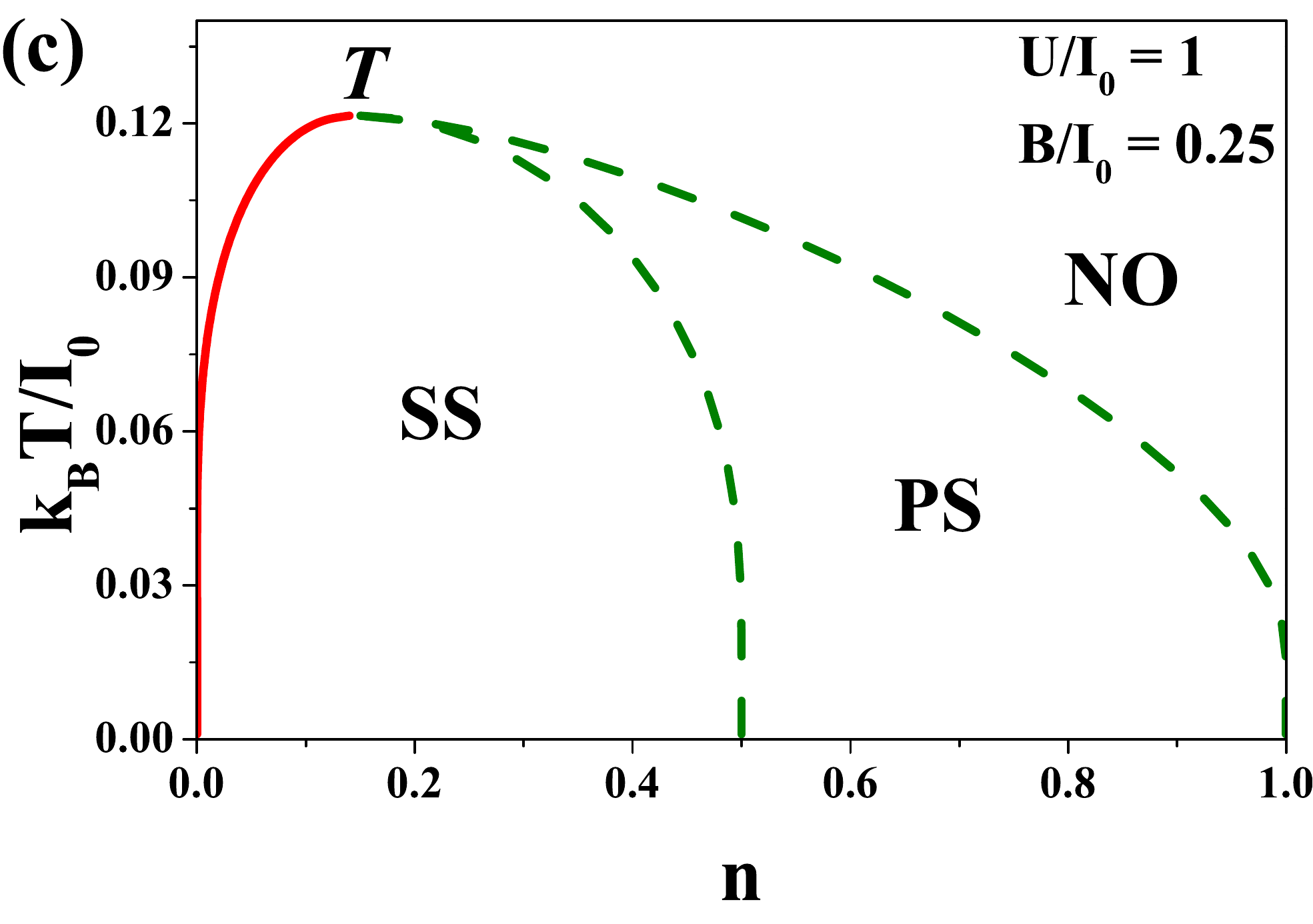}
    \caption{$k_BT/I_0$ vs. $n$ phase diagrams for  \wyr{\mbox{$U/I_0=0$}, \mbox{$B/I_0=0.99$}} (a) (corresponds to form (ii)), \wyr{\mbox{$U/I_0=0.5$}, \mbox{$B/I_0=0.5$}} (b), and \wyr{\mbox{$U/I_0=1$}, \mbox{$B/I_0=0.25$}} (c) (corresponds to form (iii)). Details in text. Solid and dashed lines indicate second order boundaries and \wyroz{(named symbolically)} ``third order''  boundaries \wyroz{(i.~e. between the PS state and homogeneous phases)}, respectively. $\mathbf{T}$ denotes the tricritical point.}
    \label{rys:PSkTvsmin}
\end{figure*}

The ground state energy of the SS phase within the VA is derived as \mbox{$f_{SS} = (1/2)Un- (1/2)|I_0| n (2-n)$} with \mbox{$|\Delta|^2 = (1/4) n (2-n)$} and \mbox{$m=0$}. For the NO phase at \mbox{$T=0$}, if $n$ is fixed, one obtains \mbox{$f_{NO} = -(1/2)B n$} (with \mbox{$2m=n$} for \mbox{$B\neq 0$} \wyr{and \mbox{$n\leq1$}}).
If $\bar{\mu}$ is fixed, one has (i) \mbox{$\omega_{NO}(\bar{\mu}) = 0$}  (\mbox{$n=0$}) and (ii) \mbox{$\omega_{NO}(\bar{\mu}) = -\bar{\mu}-U/2 - B/2$} (\mbox{$2m=n=1$}, NO -- Mott). Notice that for the NO phase the VA gives the rigorous results for the $f$ and $\omega$.
The energy of the PS state is obtained by Maxwell's construction. In the NO domain \mbox{$n_{NO}=1$} (Mott phase with \mbox{$m=1/2$} for $B\neq0$), whereas concentration in the SS domain is \mbox{$n_{SS}=1-\sqrt{(U+B)/I_0-1}$} (\mbox{$1\leq (U+B)/I_0 \leq 2$}). It corresponds to the first order \mbox{SS-NO} boundary on the $U/I_0$~vs.~$\bar{\mu}/I_0$ diagram determined by equation \mbox{$(\bar{\mu}/I_0)^2+1=U/I_0$}.
In the NO phase \mbox{$m=0$} for \mbox{$B=0$}.

It is also possible to obtain some results at \mbox{$T=0$} beyond the VA (being rigorous for $d=\infty$) in dimensions \mbox{$d=1,2,3$} by the decomposition of the eigenspace of $\hat{H}$ into sectors specified by the parity of the occupation number at each site, similarly as it has been done in \cite{KRM2012}  for $B=0$.
For $d=1$, by making use of the exact results for the ground state of the $d=1$ XY model in transverse field one can obtain the energy for the SS phase at \mbox{$T=0$}. In dimensions $1<d<+\infty$ the self-consistent random phase approximation (RPA) is a~reliable approach. It has been proven to be a~very good approximation scheme in problems  of quantum magnetism and it fully takes into account quantum fluctuations, which can be of crucial importance for the considered system for \mbox{$d\leq3$}.
The resulting diagrams are shown in figures~\ref{rys:GSdiagrams}.
For fixed $n$, the first order SS-NO transition line in the $\bar{\mu}/I_0$ vs. $(U+B)/I_0$ plane is replaced by the PS region bounded by two critical $(U+B)/I_0$ values, the lower decreasing with $n$ and the higher independent of $n$.
In finite dimensions due to quantum fluctuations the regions of the homogeneous SS phase occurrence are extended in comparison with the VA results. Moreover, at \mbox{$T=0$} in \mbox{$d=1$} only a~short-range order occurs  in contrary to the VA and RPA in \mbox{$d\geq 2$}, where the long-range order is present.

\subsection{The finite temperatures (within the VA)}\label{sec:finitetemp}

\begin{figure*}
    \centering
        \includegraphics[width=0.325\textwidth]{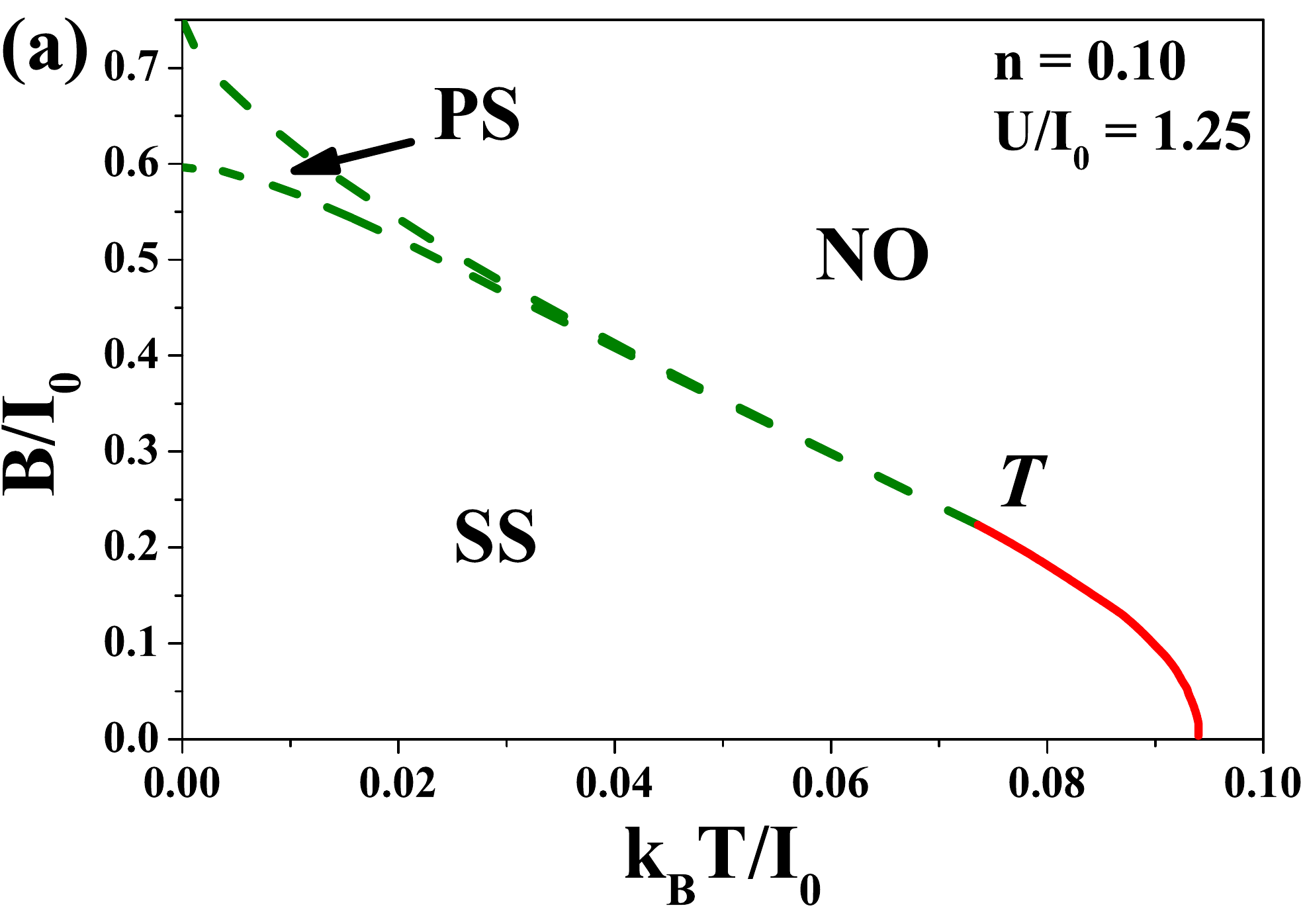}
        \includegraphics[width=0.325\textwidth]{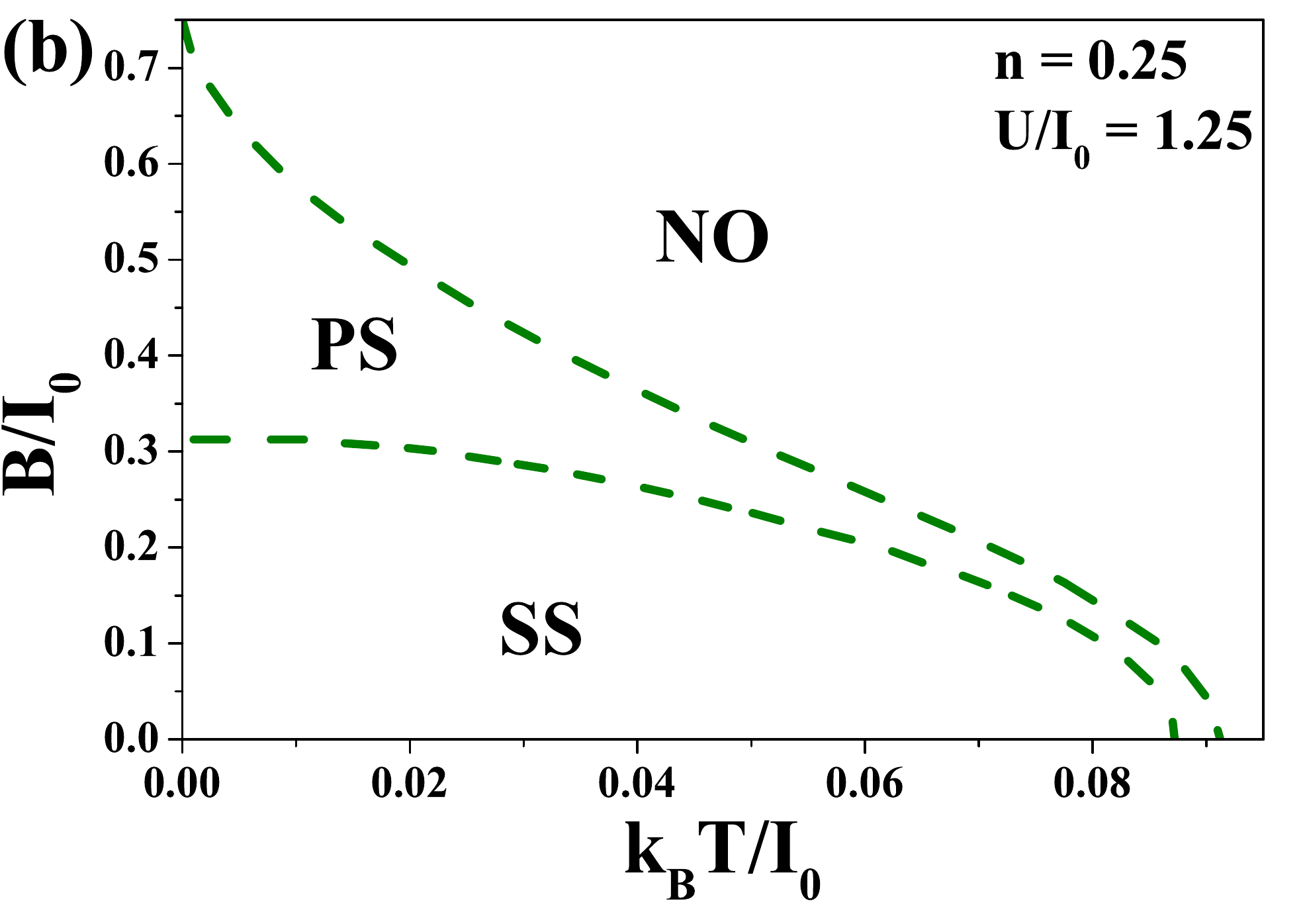}
        \includegraphics[width=0.325\textwidth]{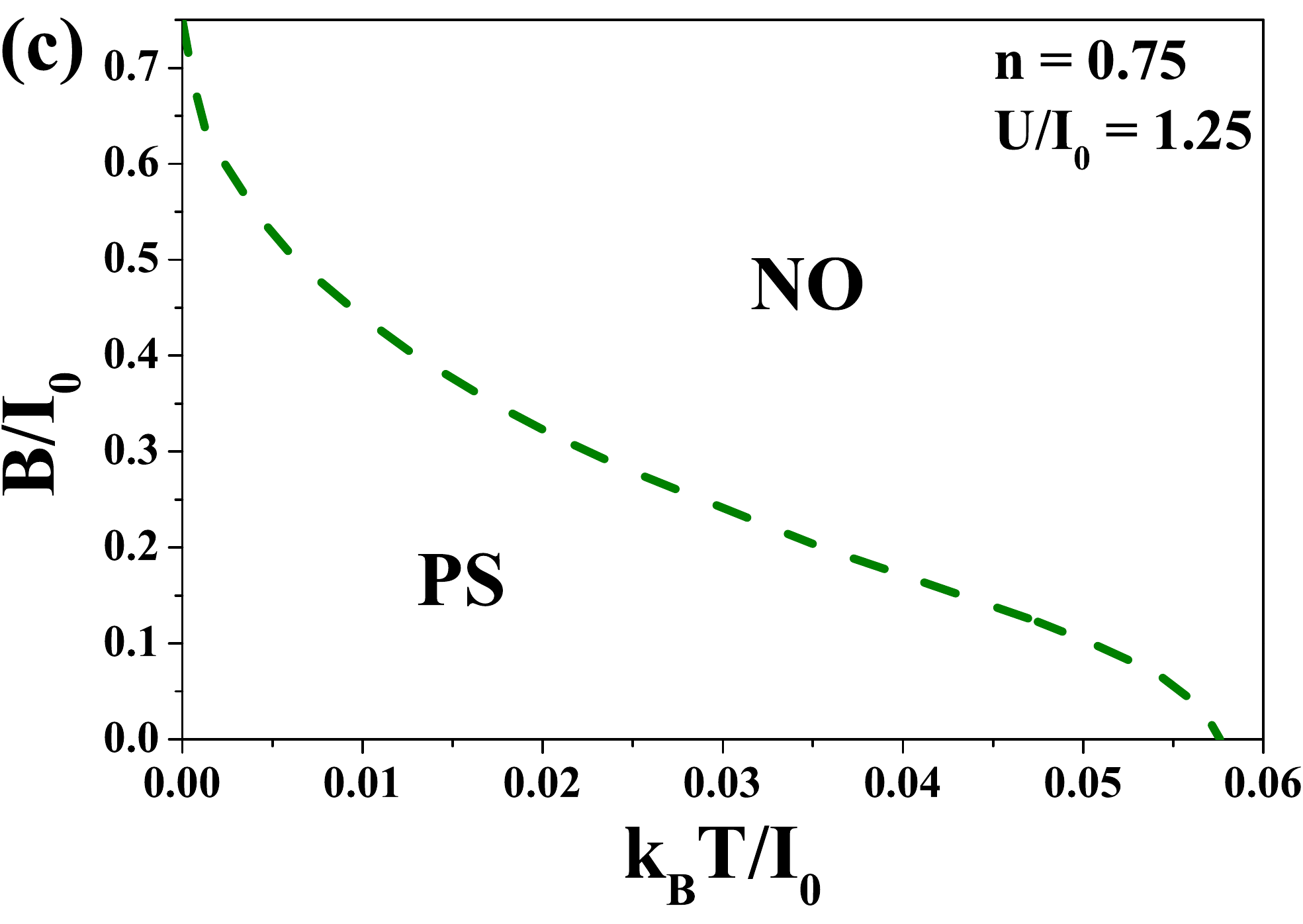}
    \caption{$B/I_0$ vs. $k_BT/I_0$ phase diagrams for  \mbox{$U/I_0=1.25$} and \mbox{$n=0.1,0.25,0.75$} (corresponding to forms (A)-(C) of the phase diagrams, respectively, defined in text). Denotations as in figure~\ref{rys:PSkTvsmin}.}
    \label{rys:PSHvskT}
\end{figure*}

\begin{figure}
    \centering
    \includegraphics[width=0.4\textwidth]{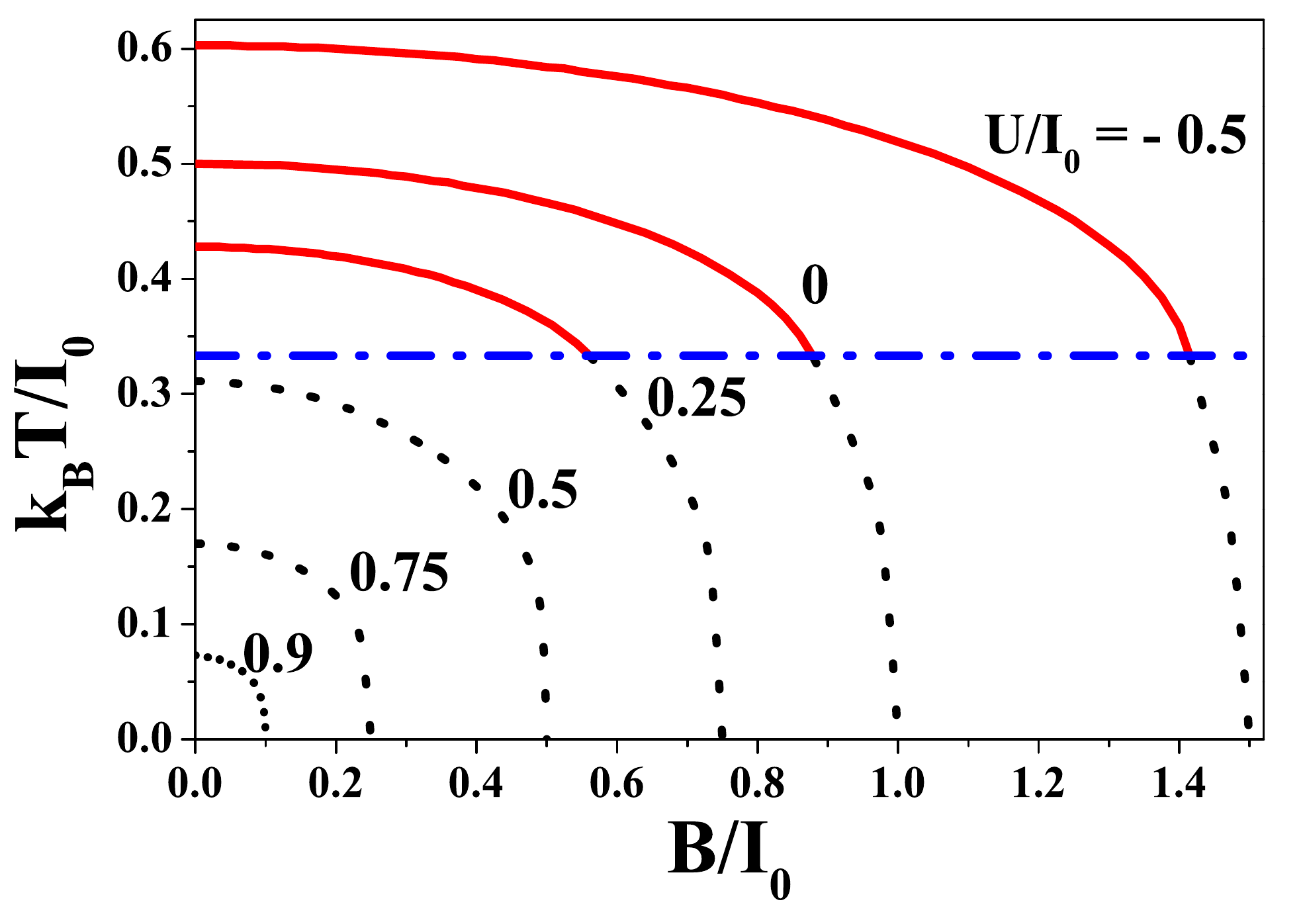}
    \caption{$k_BT/I_0$ vs. $B/I_0$  phase diagrams for \mbox{$n=1$} and various $U/I_0$ (as labelled). Denotations as in figure~\ref{rys:kTvsmi3D}.}
    \label{rys:n1}
\end{figure}

Let us discuss now the finite temperature phase diagrams  as a function of $\bar{\mu}$ (figure~\ref{rys:kTvsmi3D}).
For \mbox{$U/I_0<(2/3)\ln(2)$} the tricritical point $\mathbf{T}$, connected with a~change of the transition order, appears only for \mbox{$B>0$}. For any $|\bar{\mu}|/I_0<1$ at temperatures above $\mathbf{T}$-point the \mbox{SS-NO} transition is of second order (figure~\ref{rys:kTvsmi3D}a).
For \mbox{$(2/3)\ln(2)<U/I_0$} the line of tricritical points  starts from the \mbox{$B=0$}-plane.
If $(2/3)\ln(2)<U/I_0<1$ the SS phase with \mbox{$n=1$} can occur at finite temperatures (figure~\ref{rys:kTvsmi3D}b), whereas for \mbox{$1<U/I_0<2$} the SS phase can appear on the diagram only for \mbox{$\bar{\mu}\neq 0$} (figure~\ref{rys:kTvsmi3D}c). In the range \mbox{$2<(U+B)/I_0<+\infty$} only the NO phase is stable at any \mbox{$B\geq0$} and \mbox{$T\geq0$}. For half-filling the $\mathbf{T}$-point (if it exists) is located at \mbox{$k_BT/I_0 = 1/3$} (cf. also figure~\ref{rys:n1}).

A~few examples of $k_BT/I_0$ vs. $n$ phase diagrams are presented in figure~\ref{rys:PSkTvsmin}.
The diagrams, depending on given values of $U/I_0$ and $B/I_0$, can be  one of the following three forms only:
\begin{itemize}
    \item[(i)]
        only homogeneous phases occur on diagrams at any $T$ and the \mbox{SS-NO} transition is of second order, the transition temperatures increase with increasing $n$,
    \item[(ii)]
        the PS state appears only at \mbox{$T>0$}, e.~g. figure~\ref{rys:PSkTvsmin}a
        (the \mbox{SS-NO} first order line ends at \mbox{$T>0$} \wyr{on the $k_BT/I_0$ vs. $\bar{\mu}/I_0$ diagram for the same model parameters $U/I_0$ and $B/I_0$}),%
    \item[(iii)]
        the PS region extends from \mbox{$T=0$} (for \mbox{$1<(U+B)/I_0<2$}), e.~g. figure~\ref{rys:PSkTvsmin}c
        (the \mbox{SS-NO} first order line ends at \mbox{$T=0$} \wyr{on the $k_BT/I_0$ vs. $\bar{\mu}/I_0$ diagram}).
\end{itemize}
\wyr{Figure~\ref{rys:PSkTvsmin}b shows the limiting case between form (ii) and form (iii).}
Ranges of $B/I_0$ in which forms (i)--(iii) occur for three particular values of $U/I_0$ are collected in the upper part of Table~\ref{tab:table1}.

\begin{table}
    \caption{\label{tab:table1} Occurrence of the phase diagrams forms  ($I_0 = 1$). These forms are defined in text referring to figures~\ref{rys:PSkTvsmin} and \ref{rys:PSHvskT}.}
    \begin{ruledtabular}
        \begin{tabular}{cccc}
            Form & $U = 0$  &  $U = 0.75$ & $U = 1.25$ \\
            \hline
            (i) & $0 \leq B <0.88$ & $-$ & $ - $\\
            (ii) & $ 0.88< B <1 $ & $ 0 \leq B <0.25$  & $-$ \\
            (iii) & $ 1<B<2 $& $ 0.25< B < 1.25$ & $ 0 \leq B <0.75$ \\
            \hline
            (A) & $0<n<1$  & $0<n<0.3$ & $0<n<0.11$ \\
            (B) & $-$  & $0.3<n<1$ & $0.11<n<0.5$ \\
            (C) & $-$   & $-$ & $0.5<n<1$ \\
        \end{tabular}
     \end{ruledtabular}
\end{table}

One can distinguish three different forms of $B/I_0$ vs. $k_BT/I_0$ phase diagrams obtained for a~given fixed \mbox{$n\neq1$} (and $U/I_0$), which are shown in figure~\ref{rys:PSHvskT} (numerical values for \mbox{$U/I_0=1.25$}). For \wyr{\mbox{$U/I_0<(2/3) \ln (2)$}} only the form (A) (shown in figure~\ref{rys:PSHvskT}a) occurs for any \mbox{$0<n<1$} (cf. bottom part of Table \ref{tab:table1}). In such a~case the tricritical point $\mathbf{T}$ exist at \mbox{$T>0$}. At higher temperatures and lower fields the \mbox{SS-NO} transition is continuous (cf. also \wyr{figure~\ref{rys:kTvsmi3D}c}). At lower temperatures and higher fields the SS and NO phases are separated by the PS state. In the range \mbox{$(2/3) \ln (2)<U/I_0<1$} form (A) is realized for sufficiently small $n$, whereas for higher $n$ form (B) appears, where the regions of the SS and NO phases occurrence on the diagrams are separated by the PS state for any field and temperature (figure~\ref{rys:PSHvskT}b). For \mbox{$1<U/I_0<2$} and $n$ sufficiently close to half-filling only the PS state appears on the diagrams (absence of the homogeneous SS phase, form (C) of phase diagrams, figure~\ref{rys:PSHvskT}c).
Ranges of $n$ in which forms (A)--(C) occur for three particular values of $U/I_0$ are given in the bottom part of Table~\ref{tab:table1}.

The resulting phase diagrams for \mbox{$n=1$} and various $U/I_0$ are shown in figure~\ref{rys:n1}. In such a~case only homogeneous phases exist on the diagram. The \mbox{SS-NO} transition with increasing $T$ is of second order for \mbox{$k_BT/I_0>1/3$} (it occurs only if \mbox{$U/I_0<0.46$}) and of first order for \mbox{$k_BT/I_0<1/3$} (and any \mbox{$U/I_0<1$}).

\begin{figure*}
    \centering
        \includegraphics[width=0.325\textwidth]{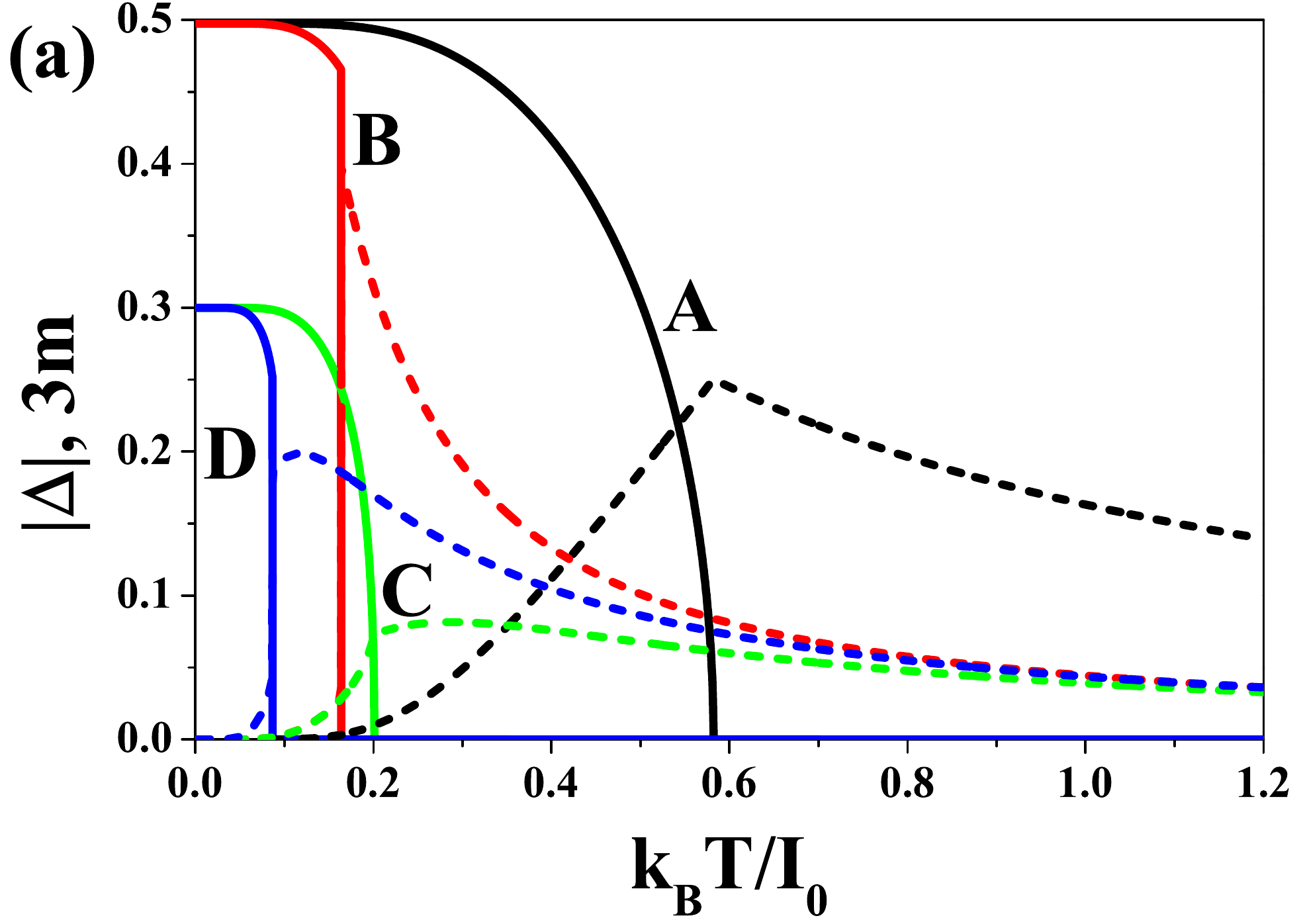}
        \includegraphics[width=0.325\textwidth]{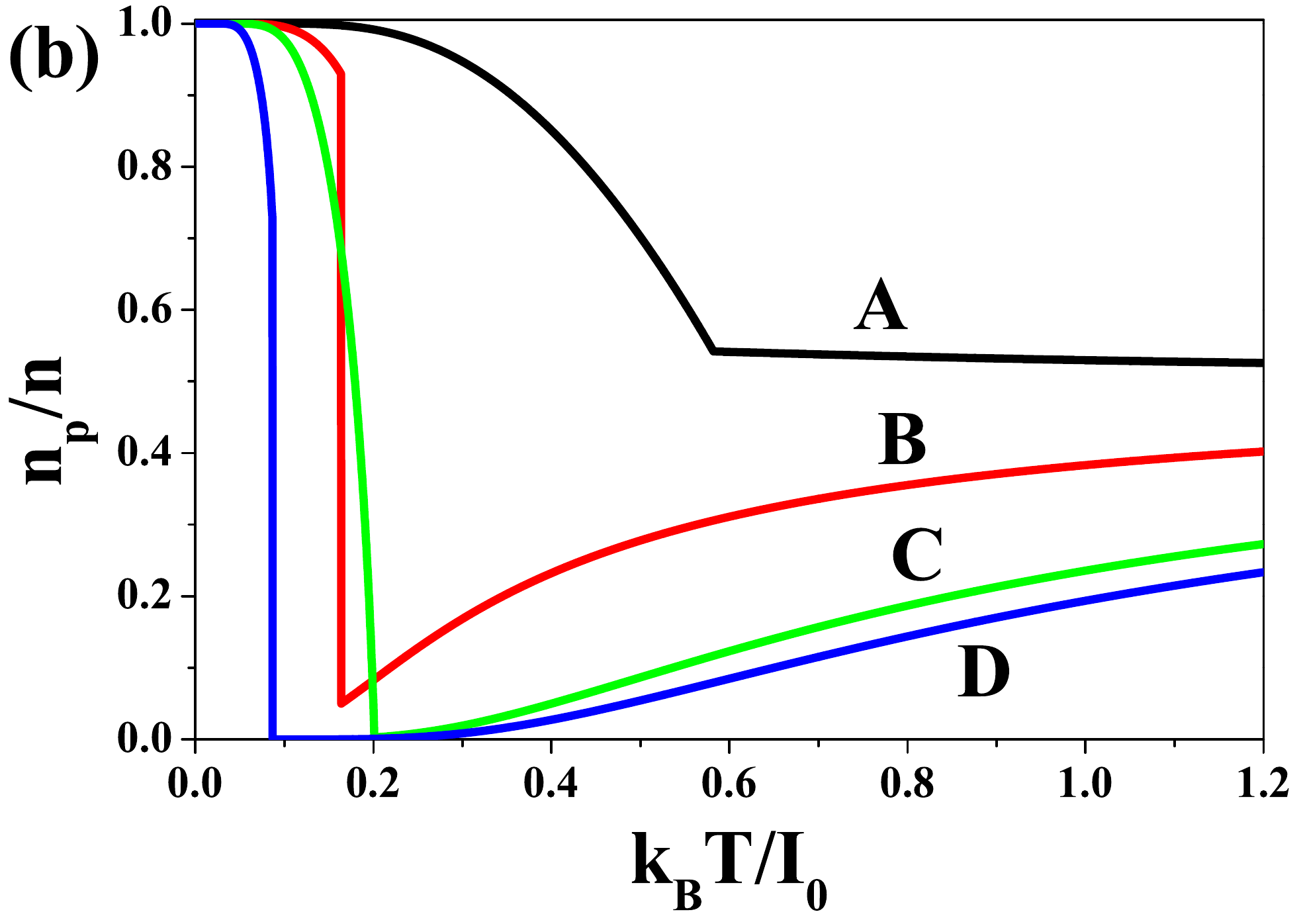}
        \includegraphics[width=0.325\textwidth]{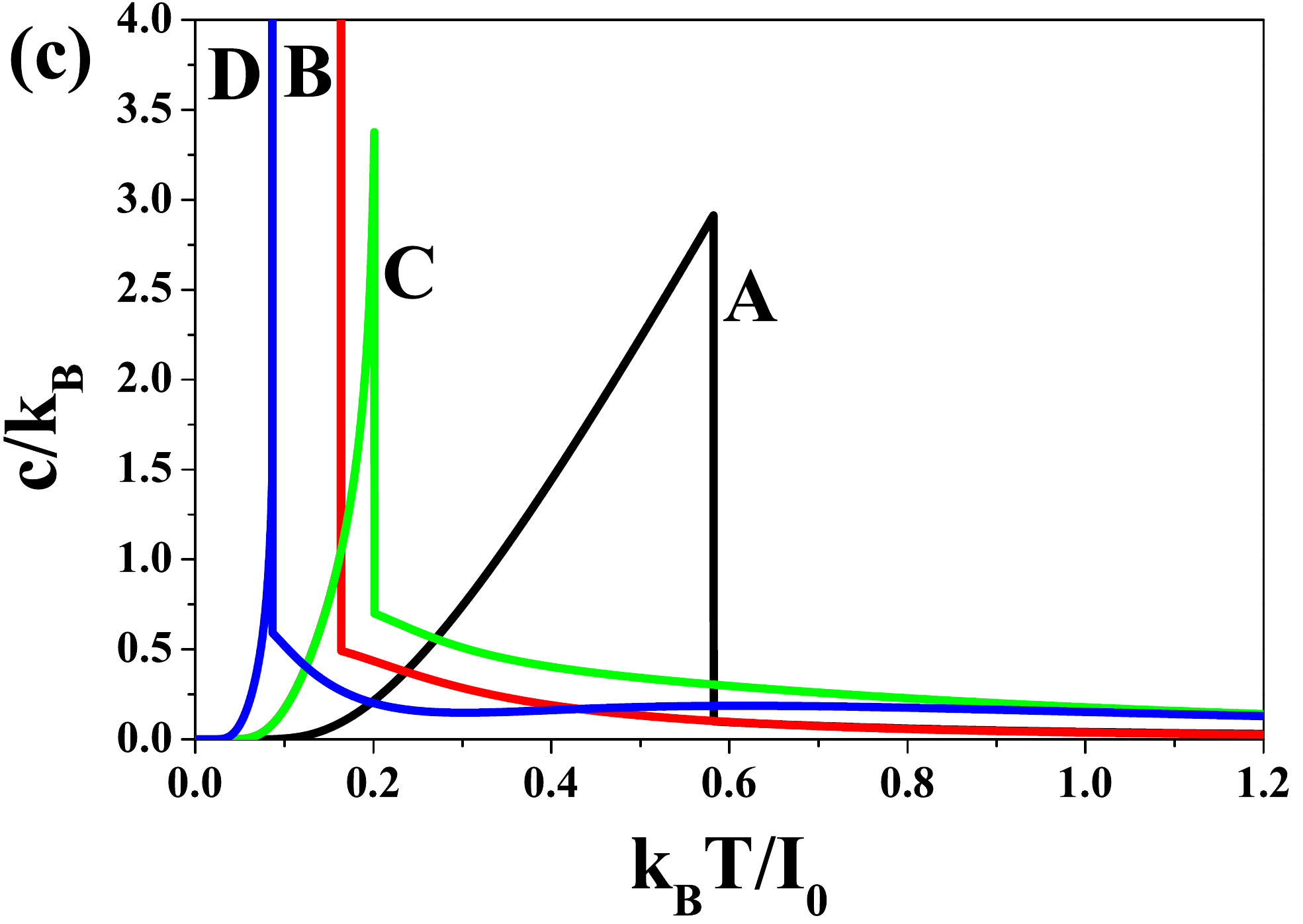}
    \caption{\wyroz{Temperature dependencies of (a)~the superconducting order parameter $|\Delta|$ (solid lines) and the magnetization $3m$ (dashed lines) (b)~the ratio $n_p/n$ and (c)~the specific heat $c/k_B$ plotted for:
    \mbox{$\bar{\mu}/I_0=-0.1$}, \mbox{$U/I_0 = -0.5$} and \mbox{$B/I_0 = 0.5$} (A),
    \mbox{$\bar{\mu}/I_0=-0.1$}, \mbox{$U/I_0 = 0.75$} and  \mbox{$B/I_0 = 0.1$} (B)
    as well as for:
    \mbox{$\bar{\mu}/I_0=-0.8$}, \mbox{$U/I_0 = 0.75$} and \mbox{$B/I_0 = 0.1$} (C),
     \mbox{$\bar{\mu}/I_0=-0.8$} \mbox{$U/I_0 = 1.25$} and \mbox{$B/I_0 = 0.1$} (D).}}
    \label{rys:HomoProp}
\end{figure*}

Concluding this section, the possible sequences of transitions with increasing temperatures and the transition orders of them for the system at fixed $n$  are listed below:
\begin{itemize}
\item[(i)]{SS$\rightarrow$NO: second order for \mbox{$n\neq1$} and second order or first order for \mbox{$n=1$},}
\item[(ii)]{PS$\rightarrow$NO: ''third order'', it can take place only for \mbox{$n\neq1$},}
\item[(iii)]{SS$\rightarrow$PS$\rightarrow$NO: both ''third order'', it can take place only for \mbox{$n\neq1$}.}
\end{itemize}

\section{Thermodynamic properties}\label{sec:termo}

\wyroz{
In this section we present several representative dependencies  of the thermodynamic characteristics for fixed model parameters. In particular, for fixed $\bar{\mu}$ (figure~\ref{rys:HomoProp}), one can single out two limiting types of thermodynamic behaviour near transition temperature $T_{SS}$: (i)~the \textit{local pair regime}
and (ii)~the \textit{pair breaking regime}.
In between, there is a~crossover between the two regimes.
\wyroz{Let us stress that in our model the single particles do coexist with pairs at finite temperatures (except \mbox{$U\rightarrow - \infty$})
and at \mbox{$T>0$} the concentration of paired electrons \wyr{$n_p$} is always smaller than $n$ and there exist finite concentration of single particles: \wyr{\mbox{$n-n_p$}} in the system. It does modify the phase transitions, which properties in the \textit{local pair regime} (\wyr{\mbox{$n_p(T_{SS})\lesssim n$}}) are different then those in the \textit{pair breaking regime}, where \wyr{$n_p(T_{SS})$} is substantially smaller than $n$ \wyr{(cf. figure~\ref{rys:HomoProp}b)}.}
}

\wyroz{
For large on-site attraction the concentration of locally paired electrons \mbox{$n_p=2D$} exhibits no sharp feature as the temperature is lowered through $T_{SS}$ \wyr{(line A in figure~\ref{rys:HomoProp}b)}. The number of non-paired electrons at $T_{SS}$ is negligible and the transition is to  the state of dynamically disordered pairs (only for \mbox{$|U|/I_0\gg1$} and \mbox{$U<0$}, the \textit{local pair regime}).
In the second limit, for on-site repulsion \mbox{$U\lesssim 2I_0$}, $n_p$ has a sharp break at $T_{SS}$ and a~substantial fraction of single particles can exists both below and above $T_{SS}$. We call this the \textit{pair breaking regime}. As temperature is lowered, the condensate grows  both from a~condensation  of pre-existing pairs and from
binding and condensation of single particles.
For small binding energies,  if \mbox{$n\ll1$} \wyr{($\bar{\mu}/I_0\approx-1$)}, there will be essentially no pre-formed pairs at $T_{SS}$ \wyr{(lines C and D)}.
}

\wyroz{Obviously a~more realistic description of coexistence of itinerant electrons and local pairs in particular real materials would be obtained using multicomponent models, especially mixed boson-fermion model (see e.~g. \cite{MRR1990,H1991,RMR1987,FL1989,MRB2005}).}

\wyroz{
Notice that non-zero value of $n_p$ does not imply that local pairs are in coherent state and even significant values of $n_p/n$ are possible in the NO phase. In the limit \mbox{$T\rightarrow+\infty$} $n_p$ increases to \mbox{$n_p/n\rightarrow 0.5$} (each of four states at a~given site can be occupied with equal probability).
The condensate density (which can be approximated as \mbox{$n_0\approx |\Delta|^2$} at least for \mbox{$n\ll1$}, \mbox{$n_0\neq D$}) vanishes for \mbox{$T\geq T_{SS}$},  but the doubly occupied states are still present above $T_{SS}$ (\mbox{$D\neq0$}).
}

\wyroz{
In figure~\ref{rys:HomoProp}a  the temperature dependencies of the superconducting order parameter $\Delta$ are presented (solid lines), where one can see clearly the discontinuous change of the order parameter (lines B and D). The other lines correspond to second order transitions. The dependencies of magnetization $m$ are denoted by dashed lines (lines for cases B and C are not distinguishable in the SS phase). One can notice that $m$ in SS phase is strongly reduced  (\mbox{$m=0$} at \mbox{$T=0$}) and increase with increasing $T$, whereas in the NO phase $m$ decreases with increasing $T$ and \mbox{$m=0$} for \mbox{$T\rightarrow+\infty$} \wyr{(for $-2\bar{\mu}>U+B$ magnetization $m$ can increase with $T$ near above $T_{SS}$, e.~g. cases C and D)}.
}

\wyroz{
Finally, let us briefly summarize the behaviour of the specific heat at constant volume \mbox{$c=-T\left(\frac{\partial^2 \omega}{\partial T^2}\right)_{\bar{\mu}}$} (figure~\ref{rys:HomoProp}c). The NO phase is characterized by the relatively broad maximum in $c$  connected with continuous changes in a~short-range electronic ordering (in higher temperatures, not shown in figure~\ref{rys:HomoProp}).  The narrow peak in $c(T)$ is associated with the first order transition, while the $\lambda$-shape behaviour is typical for the second order transition.
}

\wyroz{
The behaviour of thermodynamic parameters in PS states for the model with \mbox{$B\neq0$} is similar to that for \mbox{$B=0$}, which was widely discussed in section~5 of~\cite{KRM2012}.
}

\section{Conclusions and supplementary discussion}\label{sec:conclusions}

In this paper, we have studied the paramagnetic effects of external magnetic field on a~simple model of a~superconductor with very short coherence length (i.~e.~with the pair size being of the order of the radius of an effective lattice site) and considered the situation where the single particle mobility is much smaller than the pair mobility and can be neglected. In the model considered the pair binding energy is given by \mbox{$E_b=-U+2I_0$}, whereas the pair mobility \mbox{$\sim I$} (for \mbox{$B=0$}) and the critical magnetic field is proportional to the pair binding energy $E_b$ (at least in the low concentration limit)~\cite{RP1993}.

Let us summarize important conclusions of our work.
\begin{itemize}
    \item[(i)]
        For fixed $\mu$, if for \mbox{$B=0$} the SS--NO transition is of second order, the increasing magnetic field changes first the nature of phase transition from a~continuous to a~discontinuous type  and next it suppresses superconductivity.
    \item[(ii)]
        In definite ranges of $n$ and temperature the magnetic field stabilizes the phase separation state \mbox{SS--NO} (field-induced PS).
    \item[(iii)]
        For fixed $n$, one can distinguish four different structures of phase diagrams at \mbox{$T>0$}, as illustrated in figure~\ref{rys:PSHvskT} (\mbox{$n\neq1$}) and figure~\ref{rys:n1} (\mbox{$n=1$}).
\end{itemize}
These behaviours are associated with the presence of the tricritical point on the phase diagrams.

For \mbox{$U\rightarrow-\infty$} (states with single occupancy  are excluded and only local pairs can exists in the system) the model is equivalent with the hard-core charged boson model on the lattice~\cite{RP1993,MRK1995,BBM2002} and in such a~case there is no paramagnetic effects of $B$.
In the \emph{local pair} limit (for sufficiently strong  on-site attraction and small $B$) the charge exchange $I$ and on-site \mbox{$U<0$} cooperate and the second order \mbox{SS-NO} transition is associated with the transition to a~state of dynamically disordered local pairs. The opposite regime, i.~e. the \emph{pair breaking} limit, is realized for substantial values $U$ and $B$ (\mbox{$(U+B)/I_0\gtrapprox 1$}). In this limit the transition is determined by pair breaking excitations (driven by both $U$ and $B$, which destroy the electron pairs) and there are essentially no preformed pairs close to transition temperature (\mbox{$T\gtrsim T_c$}). In general, for fixed $U$ and $\bar{\mu}$ (\wyr{or} $n$)  the regions of ordered states occurrence are reduced by increasing $B$.

Within the VA the on-site $U$ term is treated exactly. Thus, the major conclusions of our paper concerning the evolution of the properties of the system with $U$ are reliable for arbitrary $U$.
The derived VA results are exact in the limit of infinite dimensions \mbox{$d\rightarrow+\infty$}, where the mean-field approximation treatment of the intersite interaction $I$ term becomes the rigorous one. Moreover, the VA yields exact results (in the thermodynamic limit) for $I_{ij}$ of infinite range (\mbox{$I_{ij}=(1/N)I$} for any $(i,j)$) regardless of the dimensionality (\mbox{$d<\infty$}) of the system \cite{HB1977}.
However, for short range $I_{ij}$
in finite dimensions due to quantum fluctuations the regions of the homogeneous SS phase occurrence are extended in comparison with the VA results (cf. figure~\ref{rys:GSdiagrams}). Also, in 1D-chain only a short-range order occur at any \mbox{$T\geq0$}, whereas in \mbox{$d=2$} a~long-range order can exist at \mbox{$T=0$}, while at \mbox{$T>0$} the Kosterlitz-Thouless transition is only possible.

The presence of the hopping term \mbox{$\sum_{i,j,\sigma}t_{ij}\hat{c}^+_{i\sigma}\hat{c}_{j\sigma}$}
breaks a~symmetry between the \mbox{$I>0$} (favoring SS) and \mbox{$I<0$} (favoring $\eta$S) cases.
For \mbox{$t_{ij}\neq0$} model (\ref{row:ham1}) is called the Penson--Kolb--Hubbard model \cite{HD1993}. In general for \mbox{$t_{ij}\neq0$} the phase diagrams
can involve also other ordered phases and states in addition to those obtained for Hamiltonian (\ref{row:ham1}) \cite{HD1993,RB1999,JKS2001,CR2001,DM2000,Z2005,MM2004,MC1996}, even in the ground state.
We can suppose that small but finite single electron hopping $t_{ij}$ will not qualitatively alter the phase diagrams, at least for the case \mbox{$k_BT>\sum_jt^2_{ij}/U$}. The main effect of $t_{ij}$ (for \mbox{$|U|\gg t_{ij}$}, \mbox{$U<0$})  is a~renormalization of the pair hopping term \wyr{\mbox{$I_{ij}\rightarrow I_{ij}+t^2_{ij}/|U|$}} and an introduction of an effective intersite density-density repulsion \mbox{$\sim t^2_{ij}/|U|$}. For \mbox{$U<0$} and \mbox{$I<0$} the charge density wave state can also occur \cite{KM2011,RB1999,JKS2001,PR1996}.
For \mbox{$U>0$} and both signs of $I$ the $t_{ij}$ term generates antiferromagnetic correlations (in particular for \mbox{$n=1$})  \cite{MRR1990,DL1997} competing with superconducting ones and external field $B$, and its effects can  essentially modify the phase diagrams and the properties of normal state. In such a~case it is necessary to consider also various  magnetic orderings \cite{RB1999,JKS2001,K2012}. Moreover, several phase separation states involving superconducting, charge and (or) magnetic orderings could also be stable for \mbox{$n\neq 1$}.

Although our model is (in several aspects) oversimplified, it can be useful in qualitative analysis of experimental data for real narrow-band materials with very short coherence length (exemplary systems mentioned below) and fermions on optical lattices \cite{LH2012,CX2010,ZAS2006,PLK2006,SSS2008}.
In particular, our results predict the existence of the electron phase separation (\mbox{SS--NO}) and describe its possible evolution and phase transitions with increasing $T$ and a~change of $n$ (\wyr{or} $\bar{\mu}$) in the presence of external magnetic field. Notice that the temperature dependence of the upper critical field in
unconventional superconductors
has a positive curvature in coincidence with results of figure~\ref{rys:PSHvskT}.
Obviously such a PS state is different from the Abrikosov-Shubnikov mixed-state in type-II superconductors, e.~g. no magnetic flux quantization, no vortex lattice, etc. Such ,,upper critical field'' (at \mbox{$T=0$}) is independent of $n$ and depends only on $U/I_0$ (it decreases with increasing $U/I_0$), whereas ,,lower critical field'' decreases with increasing $n$ for fixed $U/I_0$.

It is well known that there exist two distinct ways to induce pair-breaking in type-II superconductors by an applied magnetic field: orbital and spin-paramagnetic effects. The former is related to an emergence of Abrikosov vortex lines.
The spin-paramagnetic pair-breaking effect comes from the Zeeman splitting of spin singlet Cooper pairs.
The actual $H_{c2}$ of real materials is generally influenced by the both these effects.
The relative importance of the orbital and spin-paramagnetic effects can be described by the
Maki parameter $\alpha$ \cite{M1966}.
For $\alpha\ll1$ (conventional superconductors) orbital effect dominates, whereas in materials with a heavy electron mass (narrow bands) or multiple small Fermi pockets $\alpha>1$ and paramagnetic effect becomes crucial.
Some exemplary materials with $\alpha>1$
are, among others, CeCoIn$_5$, $\kappa$-(ET)$_2$Cu(NCS)$_2$, $\alpha$-(ET)$_2$NH$_4$Hg(SCN)$_4$, $\kappa$-(ET)$_2$Cu[N(CN)$_2$]Br and $\lambda$-(BETS)$_2$GaCl$_4$.
Notice also that for interacting fermions on the non-rotating optical lattices only the paramagnetic effect can occur.

In the present model the magnetic field only acts on the spin through
the Zeeman term. An interesting problem of the orbital contribution
through the pair hopping is left for future study.

The phase separation instability is specific to the short-range nature of the model.  The (unscreened) long-range Coulomb interactions prevent  the large-scale PS of charged particles \cite{PR1996} and only a~frustrated PS can occur (mesoscale, nanoscale) with the formation of various possible textures~\cite{EK1993,JCC2001,YY2004}.

The PS states involving superconductivity are shown experimentally in several systems.
For example, organic compounds exhibit the superconductor-insulator phase separations \cite{KPK2004,CSP2008,TCS2008}, whereas mesoscopic phase separation has been observed in the family of iron-pnictides \cite{PIN2009,RPC2011}.
Moreover, for special cases of  La$_2$CuO$_{4+\delta}$ and La$_{2-x}$Sr$_x$CuO$_{4+\delta}$, muon and superconducting quantum interference measurements  suggest existence of fully phase separated regions \cite{UAC2009,SFG2002,MWB2006}.
Finally, among the materials for which the local electron pairing has been either established or suggested the best candidates to exhibit the phase separation phenomena are doped barium bismutates \cite{V1988,TKP1995,PR1996}.
Recent experiments on ultracold imbalanced Fermi gas trapped in external harmonic potential serve as an alternative way to study the pure Zeeman effect on Fermi superfluidity. Several evidences \wyr{of phase separations} in such systems to state containing a paired center core and unpaired atoms outside this core have been reported \cite{ZAS2006,PLK2006,SSS2008}.

\wyroz{
It is of interest to analyze  the impact of density-density \wyr{\cite{KR2011,MRC1984}} and magnetic \wyr{\cite{KKR2010,MPS2012}} interactions on the phase diagrams of model (\ref{row:ham1}). Some results concerning the interplay of these interactions with the pair hopping term for \mbox{$B=0$}  have been presented in \cite{KRM2012,K2012}.
}

\begin{acknowledgments}
The authors wish to thank R~Micnas and T~Kostyrko for helpful discussions and a~careful reading of the manuscript.
The work has been financed by National Science Center (NCN, Poland) as a research project in years 2011-2013, grant No. DEC-2011/01/N/ST3/00413.
K.~K. would also like to thank the European Commission
and Ministry of Science and Higher Education (Poland) for
the partial financial support
from European Social Fund -- Operational Programme ``Human Capital'' -- POKL.04.01.01-00-133/09-00
-- ``\textit{Proinnowacyjne kszta\l{}cenie, kompetentna kadra, absolwenci przysz\l{}o\'sci}''
\wyr{as well as the Foundation of Adam Mickiewicz University in Pozna\'n for the support from its scholarship programme}.
\end{acknowledgments}



\end{document}